\documentclass[prl,twocolumn,showpacs,amsmath,amssymb,superscriptaddress]{revtex4}

\usepackage[dvips]{graphicx}
\usepackage{graphics}
\usepackage{dcolumn}
\usepackage{bm}
\usepackage[usenames]{color}

\begin{document}

\title{\begin{center} 1500-fold Tunneling Anisotropic Magnetoresistance in a (Ga,Mn)As stack \end{center}}

\author{C. R\"{u}ster}
\affiliation{Physikalisches Institut (EP3), Universit\"{a}t
W\"{u}rzburg, Am Hubland, D-97074 W\"{u}rzburg, Germany}

\author{C. Gould}
\affiliation{Physikalisches Institut (EP3), Universit\"{a}t
W\"{u}rzburg, Am Hubland, D-97074 W\"{u}rzburg, Germany}

\author{T. Jungwirth}
\affiliation{Institute of Physics ASCR, Cukrovarnická 10, 162 53
Praha 6, Czech Republic} \affiliation{School of Physics and
Astronomy, University of Nottingham, Nottingham NG7 2RD, UK}
 
\author{J. Sinova}
\affiliation{Department of Physics, Texas A\&M University, College Station, TX 77843-4242, USA}

\author{G. M. Schott}
\affiliation{Physikalisches Institut (EP3), Universit\"{a}t W\"{u}rzburg, Am Hubland, D-97074 W\"{u}rzburg, Germany}

\author{R. Giraud}
\affiliation{Physikalisches Institut (EP3), Universit\"{a}t W\"{u}rzburg, Am Hubland, D-97074 W\"{u}rzburg, Germany}

\author{K. Brunner}
\affiliation{Physikalisches Institut (EP3),
Universit\"{a}t W\"{u}rzburg, Am Hubland, D-97074 W\"{u}rzburg, Germany}

\author{G. Schmidt}
\affiliation{Physikalisches Institut (EP3), Universit\"{a}t W\"{u}rzburg, Am Hubland, D-97074 W\"{u}rzburg, Germany}

\author{L.W. Molenkamp}
\affiliation{Physikalisches Institut (EP3), Universit\"{a}t W\"{u}rzburg, Am Hubland, D-97074 W\"{u}rzburg, Germany}

\date{\today}

\begin{abstract}
We  report  the  discovery   of  a  super-giant tunneling
anisotropic     magnetoresistance    in    an     epitaxially    grown
(Ga,Mn)As/GaAs/(Ga,Mn)As structure.
The effect arises from a strong dependence of the electronic structure
of  ferromagnetic  semiconductors  on  the  magnetization  orientation
rather  than  from  a   parallel  or  antiparallel  alignment  of  the
contacts. The  key novel spintronics  features of this effect  are: (i)
{\it both} normal  and inverted spin-valve like signals;  (ii) a large
non-hysteretic magnetoresistance for  magnetic fields perpendicular to
the   interfaces;  (iii)   magnetization  orientations   for  extremal
resistance are,  in general,  not aligned with  the magnetic  easy and
hard axis.(iv) Enormous amplification of the  effect at low
bias and temperatures.

\end{abstract}


\maketitle  

The emerging field of  semiconductor based spintronics, which explores
the  spin  and charge  degrees  of freedom  on  an  equal footing,  is
expected to lead to  novel information technologies that will overcome
current  key obstacles  in  the microelectronics roadmap
\cite{roadmap}. A main component  needed to  realize the
full potential of  this technology is a device  with similar behavior
as  current  metal-based spin-valves  \cite{Moodera},  {\em and}  with
novel  spintronics features unattainable  in their  metal counterparts.
Previous attempts in this  direction have yielded promising spin-valve
results  \cite{Higo}  apparently mimicking  the  functionality of  the
metal devices. However, our  recent discovery of tunneling anisotropic
magnetoresitance  (TAMR)   in  a  single   (Ga,Mn)As  layer  structure
\cite{Gould} suggests   that   the   moderate
magnetoresistance (MR)  effects observed so far in  structures such as
the one in Ref.  \onlinecite{Higo} may originate from TAMR rather than
the traditional  metal tunneling MR (TMR).   If this is  the case, the
device behavior should be much  richer than for TMR, and could
offer many  new functionalities not  possible in metal  based devices.
To investigate  this hypothesis, we have fashioned  a tunnel structure
based  on the  ferromagnetic semiconductor  (Ga,Mn)As. We
report  the existence of  a huge  TAMR effect  exceeding 100~000\% in
these structures.

\begin{figure}[h!]
\includegraphics[angle=0,width=8cm]{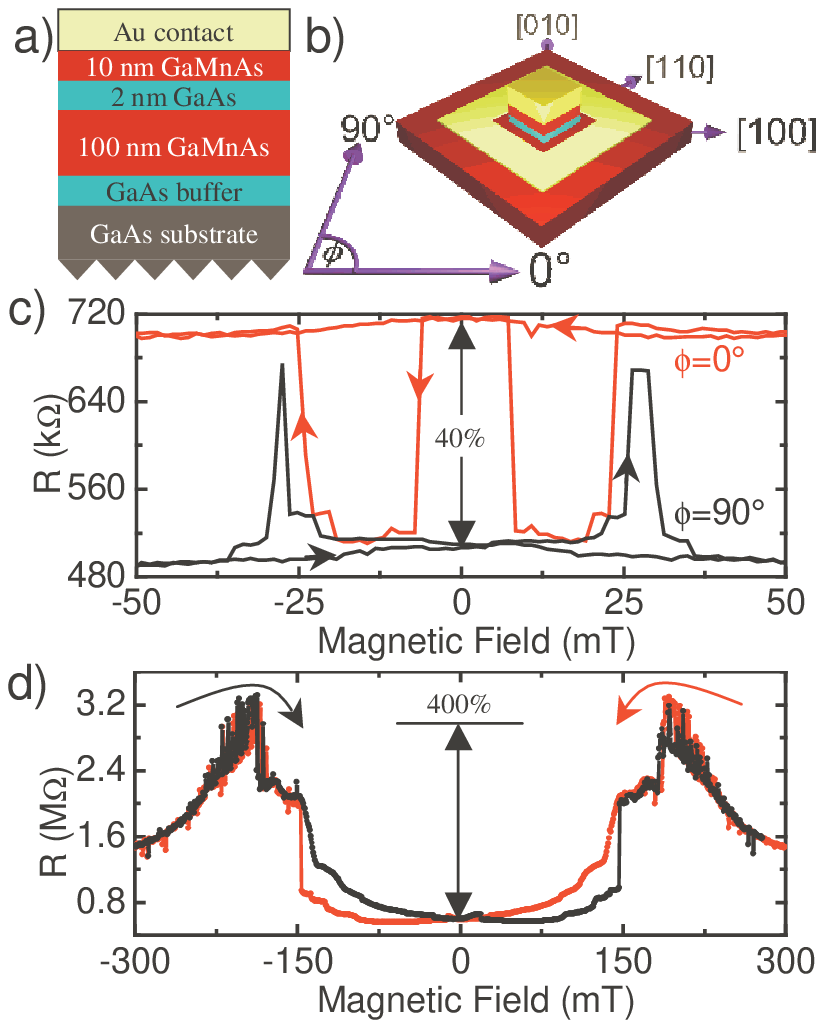}
\vspace*{-0.3cm}
\caption{a) Layer stack and b) sample layout. c) Magnetoresistance
near each in-plane easy axis showing both positive
and negative 40\% effects. d) Magnetoresistance in perpendicular field
showing a 400\% predominantly non-hysteretic signal.}
\label{figure1}
\vspace*{-.6cm}
\end{figure}

The   full    layer   structure   of   our   device, shown   in
Fig.~\ref{figure1}a, consists of a Ga$_{0.94}$Mn$_{0.06}$As
(10~nm)/GaAs   (2~nm)   /Ga$_{0.94}$Mn$_{0.06}$As   (100~nm)   trilayer
grown by low temperature molecular beam epitaxy (LT-MBE) on
a  semi-insulating  GaAs substrate  and  an undoped GaAs  buffer
layer.
The (Ga,Mn)As layers are intrinsically highly p-type due to the Mn and
have metallic  transport character. The  undoped LT-GaAs layer  on the
other hand is insulating and forms an epitaxial tunnel barrier between
the   two   ferromagnetic   layers.   The   ferromagnetic   transition
temperature of the (Ga,Mn)As is $\sim$ 65 K.

A    schematic    of    the    patterned   device    is    shown    in
Fig.~\ref{figure1}b. Using   optical   lithography   with   positive
photoresist followed  by metal evaporation, lift-off,  and wet etching
the heterostructure  was patterned  into a square  mesa with  sides of
100~$\mu$m and  a surrounding electrical back contact. Contact to the
square mesa is established by  an in-situ Ti/Au deposition, whereas the
back contact  is made  after patterning by  depositing Ti/Au  onto the
lower (Ga,Mn)As  layer. Two-probe MR measurements  are then performed
for current flowing vertically through the layer stack. Since the bulk
resistivity      of     the      (Ga,Mn)As is     only
$\sim$$10^{-3}$~$\Omega$cm, the device  resistance is dominated by the
tunnel barrier. Identically patterned control samples without a tunnel
barrier  have a resistance  of $10$~$\Omega$,  proving that  any bulk
(Ga,Mn)As MR is fully negligible.

Transport measurements  were carried out in  a magnetocryostat fitted
with  a  variable  temperature  insert  and  three  sets  of  mutually
orthogonal Helmholtz coils allowing the application of magnetic fields
\textit{$\bf{H}$} of up to 300~mT in any direction. Fields applied in
the plane of the layers are denoted by their angle $\phi$ with respect
to  the  [100]  crystallographic  direction. Two  different  types  of
\textit{$\bf{H}$} scans will be  presented: MR-scans, which consist of
saturating the sample magnetization in a negative magnetic field along
a given direction  and then measuring the resistance  of the device as
$|$\textit{$\bf{H}$}$|$  is  swept  to  positive saturation  and  back
again; and $\phi$-scans, where the resistance is measured for constant
$|$\textit{$\bf{H}$}$|$ while sweeping $\phi$.

Fig.~\ref{figure1}c  shows  MR-scans  taken  with a bias voltage V$_{B}$ = 10~mV at a temperature  T = 4.2~K along $\phi$ = 0$^{\circ}$
(red) and 90$^{\circ}$ (black), near  the two cubic magnetic easy axes
in  the  (Ga,Mn)As  ([100]  and  [010] respectively)  as  verified  by
SQUID. At  low $|$\textit{$\bf{H}$}$|$  after crossing zero  in either
sweep    directions,   \textit{$\bf{M}$}    abruptly    reverses   its
direction.  This  manifests itself  in  transport  as a  discontinuous
change  in  resistance  leading  to  a ~40\%  spin-valve  signal.  The
measurement   along   90$^{\circ}$   appears   similar   to   previous
observations \cite{Higo}, and could easily be mistaken for traditional
TMR.    However,   the    remarkable   sign    change    observed   at
$\phi$=0$^{\circ}$  points to a  different origin  of the  effect, and
strongly  suggests  an  interpretation   in  line  with  our  previous
observations of TAMR in single-ferromagnet devices \cite{Gould}.

As we apply $|$\textit{$\bf{H}$}$|$ at  other angles in the plane, the
amplitude of the effect remains constant whereas the position and sign
of the  sharp switching events  displays a strong  angular dependence,
with an underlying symmetry  consistent with the  one reported  for a
single  magnetic  layer  device  \cite{Gould}.  Neglecting  some  fine
structure in  the shape of  the peaks, the  relatively straightforward
picture of the two step magnetization reversal reported in \cite{Gould} accounts for this low  $|$\textit{$\bf{H}$}$|$ symmetry. It comes from a combination of the magnetic anisotropy of the
(Ga,Mn)As, which is principally cubic with a small in-plane uniaxial contribution, and the fact that magnetic  reversal  takes place  via 90$^{\circ}$ domain  wall nucleation  and propagation. At  low fields,
rather than  a coherent rotation of the  magnetization, the dominating
reversal  mechanism consists of  the magnetization  switching abruptly
whenever the energy gain by doing so is greater than the energy needed
to nucleate/propagate a domain wall. This leads to the symmetry of the
magnetization reversal  consistent with previous  studies in epitaxial
Fe layers \cite{Cowburn}.

Another observation distinguishing our effect from TMR is
a strong MR signal observed when \textit{$\bf{H}$} is
applied perpendicular to the plane of the sample, i.e., along
the magnetic hard axis. 
Fig.~\ref{figure1}d shows such
a MR-scan at T = 4.2~K and V$_B$ = 5~mV. 
The TAMR in
Fig.~\ref{figure1}d is $\sim 400\%$, much larger than for
\textit{$\bf{H}$} in-plane under similar conditions. As explained in the theory
section,
we attribute this to a significant 
growth direction strain in our (Ga,Mn)As layers that induces
a large anisotropy between the [001] and [100] (or [010]) directions, compared to
the relatively weak in-plane uniaxial anisotropy mentioned above.
Note
also that the perpendicular TAMR is no longer hysteretic, but occurs
on both sides of \textit{$\bf{H}$}=0, 
indicating that it must be related to the absolute rather than the
relative orientations of the ferromagnetic layers.

\begin{figure}[h]
\vspace*{-0.1cm}
\center{\includegraphics[angle=0,width=8.4cm]{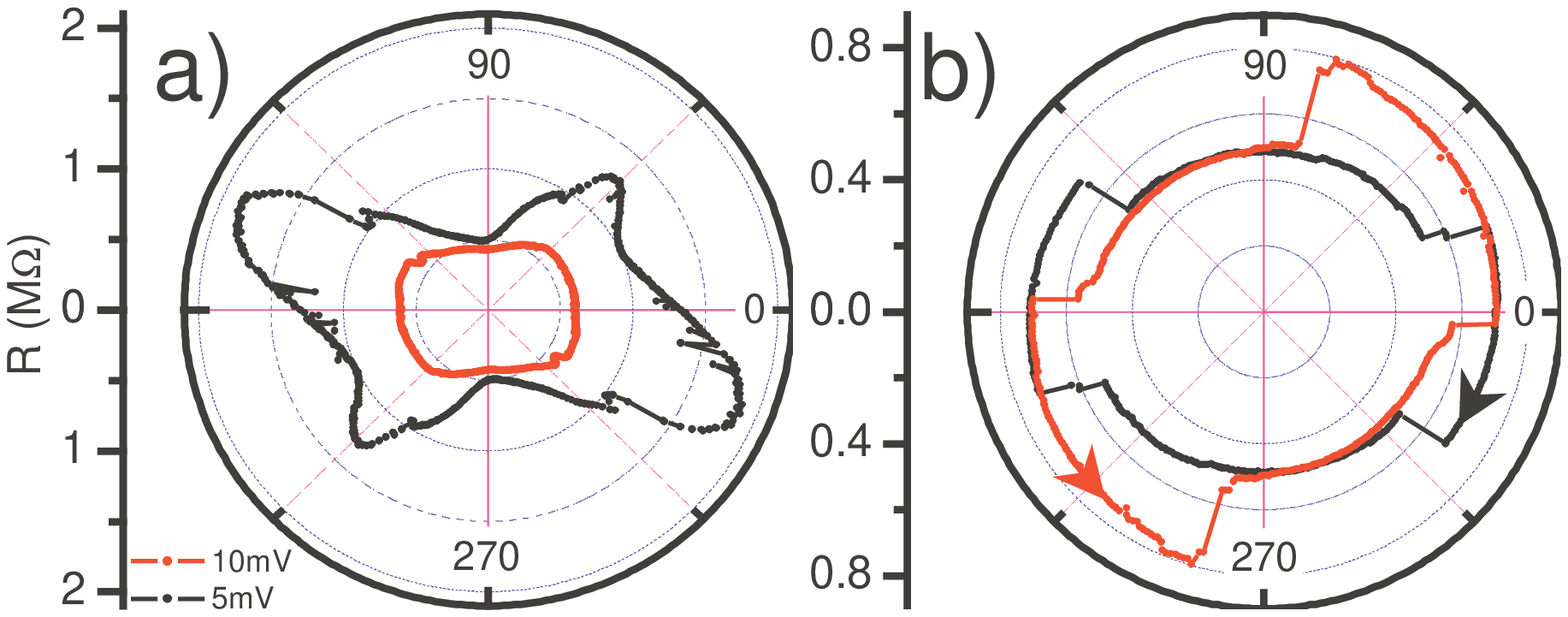}}
\vspace*{-9cm}
\caption{$\phi$-scans  at 4.2~K  a) in  a  saturation magnetic  field
$|$\textit{$\bf{H}$}$|$ = 300~mT,  and b) $V_{B}$ = 5 mVat $|$\textit{$\bf{H}$}$|$ =
25~mT, just sufficient to switch M between easy axes.}
\label{figure2}
\end{figure}
\vspace*{-0.2cm}

The TAMR is further  evidenced through $\phi$-scans at higher magnetic
fields. As can already  be seen in Fig.~\ref{figure1}c, the resistance
at saturation is dependent  on the direction of magnetization, varying
from  480 to  700~k$\Omega$  as \textit{$\bf{M}$}  changes from  along
[010]  to  [100]. This  is  a  unique  characteristic of  our  device,
since  in contrast to  regular TMR which depends  only on
the relative orientation of  \textit{$\bf{M}$} in the two layers, our
sample is sensitive to the absolute directions of \textit{$\bf{M}$} in
the layers. Therefore it can act as a sensor of the absolute direction
of   \textit{$\bf{H}$}.   This characteristic  is exhibited in the 5 mV, $|$\textit{$\bf{H}$}$|$ = 300~mT $\phi$-scan of Fig.~\ref{figure2}a.   The  measurement  is
identical  for  clockwise  or  counter clockwise  $\phi$  sweeps  as
$|$\textit{$\bf{H}$}$|$ is sufficiently     large     to     saturate
\textit{$\bf{M}$}   such  that   \textit{$\bf{M}\propto\bf{H}$}.   The
resistance  changes  by  more   than  250\%  between  its  minimum  at
90$^{\circ}$ and its maximum  at 165$^{\circ}$.  The fact
that the  maximum resistance does not  lie along the  main crystal and
anisotropy axes will be addressed below.

We  turn now our  attention to  Fig.~\ref{figure2}b, showing  that the
$\phi$-scan  changes dramatically  with  $|$\textit{$\bf{H}$}$|$. Here
$|$\textit{$\bf{H}$}$|$=  25~mT was  chosen to  be slightly  above the
biggest  coercive  field  in  the  sample  such  that  it  can  switch
\textit{$\bf{M}$}  from  one easy  axis  to  the  other as  $\phi$  is
swept.  The  sample  was  prepared  in a  known  state  by  saturating
\textit{$\bf{M}$}  along  the uniaxial  easy  axis (90$^{\circ}$)  and
$|$\textit{$\bf{H}$}$|$ was  then lowered to 25~mT.  After this $\phi$
was swept in the clockwise  (black) or counter clockwise (red
curve) direction.

The main features of  the data are $\sim 40$\% jumps in
the resistance between the 500  and 700 k$\Omega$ levels. These can be
understood rather simply by noting that at $\phi=90^{\circ}$ the sample
is in  a low resistance state associated  with \textit{$\bf{M}$} being
along the [010] easy axis. As $\phi$ is swept nearer to the [100] easy
axis,  \textit{$\bf{M}$}  will eventually  switch  to this  direction,
corresponding  to  a  high  resistance  state due  to  the  additional
uniaxial  field  that  breaks  the  in-plane  cubic  symmetry  in  the
(Ga,Mn)As  layers.  The  curves must  be different  for the  two sweep
directions as  they should have approximate mirror  symmetry about the
easy  axis. The  deviations from  this symmetry may be  attributed to
non-uniform strain distributions.

A few additional levels are seen in the data on the edges of the large
switching  events. These  intermediate states  can be  explained  in a
straightforward way.  By design, the magnetic anisotropies  of the two
layers  are   not  identical   as  different  strain   conditions  and
thicknesses    create   different    coercive    fields.   Thus,    as
\textit{$\bf{H}$} is rotated, the layers do not switch simultaneously,
but  the softer  layer switches  earlier. This  creates configurations
where  the magnetizations  of the  two layers  are not  collinear, but
perpendicular  to  each other.  As  a  control  experiment, a  similar
$\phi$-scan at $|$\textit{$\bf{H}$}$|$ =  15~mT was made. As expected,
since  15~mT  is  just  below  the  smallest  coercive  field  in  the
structure, no switching  took place, and the resistance  of the sample
remained  constant at  its lowest  value. This  behavior suggests
design  perspectives  for  spin  valves programmable  in
rotating magnetic fields above  a certain threshold magnitude, but not
below.

The   data   in   Figs.~\ref{figure1}  and   \ref{figure2}   establish
unambiguously the TAMR nature of the measured effect.  Anisotropies in
the   (Ga,Mn)As   density   of    states   (DOS)   with   respect   to
\textit{$\bf{M}$},   which   result   from  the   strong
spin-orbit coupling in  the ferromagnetic semiconductor valance band,
are large  enough to explain the effect  \cite{Gould}. Consistently, a
sizable anisotropy of the MR of (Ga,Mn)As-based tunnelling structures,
comparing the magnetization parallel or perpendicular to the tunneling
current, was  found in an  independent theoretical work  employing the
Landauer   transport  formalism   \cite{Brey}.   The  DOS   anisotropy
calculations in Ref.\onlinecite{Gould}, based on the kinetic-exchange
model coupling of  valence band holes and polarized  Mn local moments,
have  explained the change  of the  sign of  the TAMR  spin-valve like
signal  with  field  angle   and  temperature,  and  predicted  strong
enhancement of the effect  in epitaxial tunnel junctions characterized
by  a larger  degree of  in-plane momentum  conservation.  Our present
study  confirms  this prediction  and  some  of  the new  experimental
features of  the (Ga,Mn)As/GaAs/(Ga,Mn)As TAMR can  also be understood
based on the DOS anisotropy.

The  relative DOS  anisotropies,  $\Delta$DOS$_{int}$/DOS$_{int}$, for
the DOS at the Fermi energy $E_F$, integrated over an assumed range of
momenta  k$_z$ along  the  tunnelling direction  and  summed over  the
occupied spin-split valence  bands, are plotted in Fig.  3 for several
in-plane  magnetization orientations.  Note  that states
at  $E_F$ with  the largest  k$_z$ are  expected to  have  the largest
tunnelling  probability  when  in-plane  momentum  is  conserved  and,
therefore,  increasing   the  range  of  k$_z$   contributing  to  the
tunnelling   DOS  corresponds  to   relaxing  the   in-plane  momentum
conservation condition,  or increasing the tunnel barrier
transparency.  Theoretical curves in  panels (a),  (b), and  (c) were
obtained  for  hole  densities  0.1,  0.5, and  1  $\times$  10$^{20}$
cm$^{-3}$,  respectively.  Although the  bulk  hole  densities in  our
(Ga,Mn)As  layers  are of  order  10$^{20}$  cm$^{-3}$, a  significant
depletion is expected in  the region near the (Ga,Mn)As/GaAs interface
with hole  concentration closer  to those in  panels (b) or  (a).  The
substitutional Mn$_{\rm  Ga}$ concentration  of 4\% considered  in the
calculations is  consistent with  the nominal total  Mn doping  in our
(Ga,Mn)As  layers. A  uniaxial strain  along [010]  was  introduced to
model the broken in-plane cubic symmetry in the (Ga,Mn)As.

\begin{figure}[h]
\vspace*{0cm}
\includegraphics[angle=0,width=8.0cm]{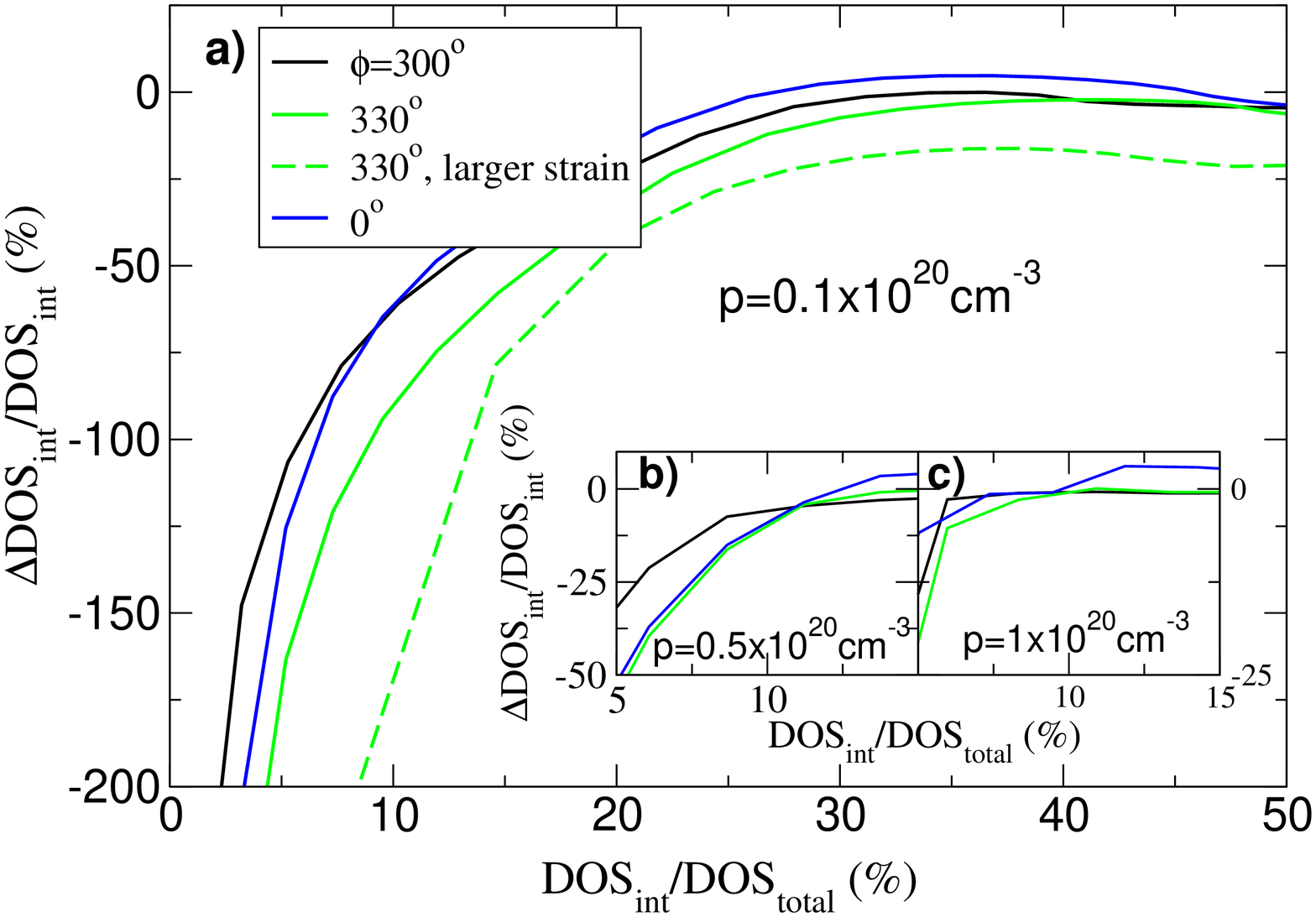}
\vspace*{-0.5cm}
\caption{Theoretical diagrams obtained for hole densities 0.1 (a), 0.5 (b), 
and 1 $\times$ 10$^{20}$ cm$^{-3}$ (c) showing the relative  
difference between the integrated DOS at the Fermi energy for
M along $\phi=270^{\circ}$) and three different angles $\phi$.
The x-axis represents the integrated DOS at the Fermi energy
that is assumed to contribute to tunneling, relative to the total
DOS at the Fermi energy. 
Solid (dashed) lines were obatined for a uniaxial strain
along [010] direction of 0.2\% (0.4\%).}
\label{figure3}
\end{figure}

Fig.~\ref{figure3} demonstrates that the magnitude of the
DOS$_{int}$  anisotropy  as  well  as the  magnetization  orientations
corresponding to extremal tunnelling  DOS have a complex dependence on
the magnetic tunnel junction parameters. Data in panel (a), e.g., show
DOS$_{int}$ anisotropies       exceeding        100\%       for
DOS$_{int}$/DOS$_{total}\sim$ 10\%.   (DOS$_{total}$ denotes the total
DOS  at  $E_{F}$.)   Here  the  minimum DOS$_{int}$  is  for
\textit{$\bf{M}$} at $\phi=270^{\circ}$ while the maximum DOS$_{int}$
is  at $\phi=330^{\circ}$,  i.e., off  the main  crystal  and magnetic
anisotropy axis.  The result provides an explanation for the distorted
cubic symmetry observed in  the experimental in-plane angle dependence
of  the TAMR  (see  Fig.~\ref{figure2}a). The  enhanced DOS  anisotropy
shown by  the dashed line  in the main  panel, which was  obtained for
larger strain  value (larger magnetic anisotropy),  is consistent with
the experimentally observed enhancement of the TAMR when magnetization
is switched between the easy- and hard-axis (see Fig.~\ref{figure1}).
We emphasize, however, that the theoretical data in Fig.~\ref{figure3}
are  only illustrative;  a  more quantitative  comparison between  the
experiment  and theory  requires  a detailed  characterization of  the
experimental tunnel  junction parameters and  a systematic theoretical
analysis of  the TAMR and  TMR contributions to the  hole transmission
coefficients, which  is beyond the  scope of this  short communication
and will be published elsewhere.

\begin{figure}[h]
\hspace*{-0cm}
\includegraphics[angle=0,width=8.4cm]{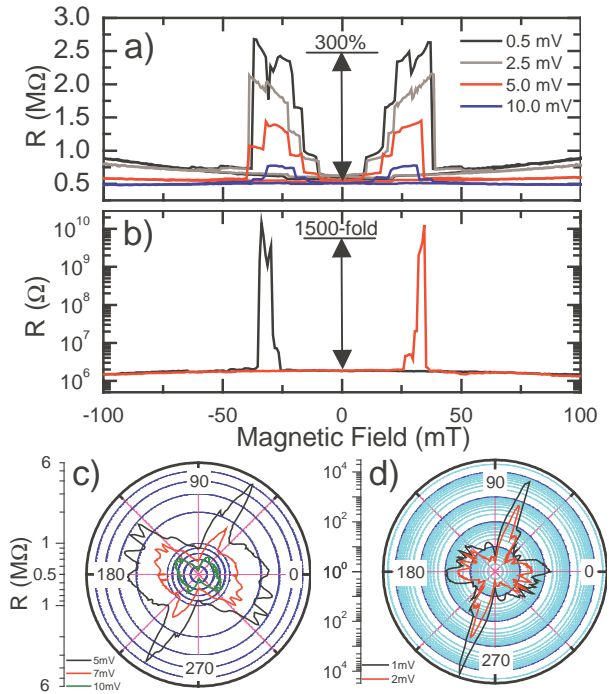}
\vspace*{-3cm}
\caption{Amplification  of   the  effect  at  low   bias  voltage  and
temperatures. a) TAMR along $\phi$ = 65$^{\circ}$ at 4.2~K for various
bias voltages. b) Super-giant  TAMR at 1.7~K and 1~mV  bias. c) and d)
$\phi$  at  various  bias at  1.7~K  showing  that  at low  bias  and
temperatures, TAMR probes the  detailed anisotropies of the density of
states.}  \label{figure4}
\end{figure}
\vspace*{0cm}

Another  prominent characteristic  of our  device is  the  very strong
V$_B$ dependence  of the signal displayed  in Fig.~\ref{figure4}a. The
various curves show the MR along $\phi$ = 65$^{\circ}$ at \textit{T} =
4.2~K  with  V$_B$  ranging  from  500~$\mu$V up  to  10~mV.  The  low
resistance state has a relatively  small variation of $\sim$ 20\% with
decreasing bias.  In contrast, the high resistance  state increases by
more than 250\%. The amplitude of the TAMR effect is also very sensitive to
$T$, as  shown in  Fig. 4b. Here we show a V$_B$=1~mV  curve at 1.7~K  where the  effect has grown to 150~000\%.  Indeed, this  is
merely  a lower  limit corresponding  to  the detection  limit of  our
current  amplifier. Although  the  amplitude of  the effect  increases
dramatically  at  low   \textit{V$_B$}  and  \textit{T},  the  general
symmetry remains unchanged indicating that the origin of the effect is
unchanged, but that it is amplified by an additional mechanism.

\vspace{-0.05cm}
This super-giant amplification of the TAMR can be understood as a manifestation
of a well known zero bias anomaly \cite{Lee} in tunneling from a
dirty metal which appears due to the opening of an Efros-Shklovskii
gap \cite{Efros} at $E_F$ when crossing the metal-insulator
transition. Indeed, such an effect should be observed in our device
given the short (Ga,Mn)As mean free path of a few \AA~\ which limits the
injector  region to a very thin layer near the barrier. Depletion near
the barrier must therefore cause a lower carrier density in the
injector region than in the bulk of the (Ga,Mn)As slab. The injector
will therefore be much closer to the metal-insulator transition than a
typical (Ga,Mn)As layer. Moreover, we already know that the DOS
changes with \textit{$\bf{M}$}. Therefore, when we perform
experiments at  low \textit{V$_B$} and \textit{T}, the effective DOS
participating in the tunneling can be brought through the
metal-insulator transition with reorientation of
\textit{$\bf{M}$}, leading to a large amplification of the
TAMR effect. A further indication that the  Efros-Shklovskii gap is the dominant enhancing 
mechanism is that the amplification of the effect as $T$ changes from 4.2 to
1.7 K is strong for low bias voltage (1 mV), but disappears at higher voltages (10 mV), 
consistent with tunneling experiments near the metal-insulator transition of Si:B \cite{Lee}.
Other  possible mechanisms for the enhancement  of the TAMR,
such as disorder and impurity mediated tunneling, may also play a role
and  should  not be  summarily  dismissed.   Further experimental  and
theoretical study  of the problem  is needed before  the amplification
mechanism can be claimed to be fully understood.

Finally, in Fig.~\ref{figure4}c and d we present $\phi$-scans at 1.7~K
for various  V$_B$, which demonstrate another important caracteristic of our
device which is that it acts as a detector for the anisotropies in the
DOS  of the  (Ga,Mn)As layer.  Fig.~\ref{figure4}c already  shows some
fine structure,  which becomes  much more pronounced  at  lower the
bias.  This is  to be expected since we start  detecting fine
structure in the anisotropy of  the DOS, which should be complex given
that  the opening  of the  gap  should develop  differently for  the
various bands which have different effective masses.

In   summary,    we   have   observed    a super-giant TAMR   effect    in  a
(Ga,Mn)As/GaAs/(Ga,Mn)As tunnel  structure which can be of  order of a
few  hundred \%  at 4K,  and can  be amplified  to 150~000\%  at lower
temperatures. The behavior of the structure not only mimics normal TMR
when the field is applied along the [010] direction, but also exhibits
new functionalities such  as a sensitivity to not  only the amplitude,
but also  to the direction  of an applied  magnetic field.
While many  of the experimental features  of this novel  effect can be
understood  through the one-particle  tunneling DOS  anisotropies with
respect to  the magnetization orientation,  the dramatic amplification
at low biases and temperatures poses new challenging questions for the
theory  of tunneling  transport in  disordered  interacting electronic
systems with strong spin-orbit interaction.

The authors thank A.H. MacDonald and M. Flatt\'{e} for useful discussions and V. Hock
for sample preparation, and acknowledge financial support from the EU (SPINOSA), the German BMBF (13N8284) and DFG (BR1960/2-2), the Czech Republic GACR (202/02/0912), and the DARPA SpinS
program. R.G. thanks the Humboldt foundation for financial support.

\end{document}